\documentclass[a4paper,11pt,authoryear]{article}
\pdfoutput=1 

\usepackage{jcappub} 

\usepackage[natbib=true,
    style=numeric-comp,
    sorting=none]{biblatex}
\usepackage[T1]{fontenc}
\usepackage{amssymb}

\title{Adiabatic and isocurvature perturbations in extended theories with kinetic couplings}

\author[]{Mariaveronica De Angelis and Carsten van de Bruck}

\affiliation[]{Consortium for Fundamental Physics, School of Mathematics and Statistics, University of Sheffield, Hounsfield Road, Sheffield S3 7RH, United Kingdom}

\emailAdd{mdeangelis1@sheffield.ac.uk}
\emailAdd{c.vandebruck@sheffield.ac.uk}

\abstract{The scalar field sector in low--energy effective field theories motivated by string theory often contains several scalar fields, some of which possess non--standard kinetic terms. In this paper we study theories with two scalar fields, in which one of the fields has a non--canonical kinetic term. The kinetic coupling is allowed to depend on both fields, going beyond the work in the literature, which usually considers the case of the coupling to depend on the other field only. Our aim is to study adiabatic and isocurvature perturbations in these extended theories. Our results show that the evolution equation for the curvature perturbation does not change when allowing the coupling to depend on both fields, while the effective mass of the entropy perturbation changes.  We find expressions for the spectral index and its running at horizon crossing and at the end of inflation. We apply the formalism and study three phenomenological models, with different kinetic couplings. 
}

\begin{document}
\maketitle
\flushbottom

\section{Introduction}
Inflation in the very early universe is the most influential idea about the origin of cosmic structures \cite{Guth:1980zm,Starobinsky:1980te,Linde:1983gd,Albrecht:1982wi}, see \cite{Baumann:2009ds} for a review. According to this theory, the universe underwent a period of accelerated expansion in the very early universe. It solves some basic paradoxes of the standard hot Big Bang model. But despite this, the inflationary scenario has yet still to be embedded into a fundamental theory. There are many attempts to realise a period of inflation within a more fundamental theory, such as supergravity or string theory (to name a few references, see e.g. \cite{Kachru:2003sx,Firouzjahi:2003zy,Baumann:2007ah,McAllister:2014mpa,Burgess:2011fa,Burgess:2013sla,Baumann:2014nda}), but in many of these inflationary models there is the need for a certain amount of fine-tuning of the model parameters (especially concerning the flatness of the scalar field potential). Recently, the swampland program within string theory put a significant hurdle on inflationary model building, in that according to the conjectures, scalar field potentials cannot be arbitrarily flat and the field excursion has to be sub--Planckian \cite{Obied:2018sgi,Agrawal:2018own}. Inflation also does not address the initial singularity \cite{Borde:2001nh}. 

Despite these theoretical hurdles, the predictions of inflation are in excellent agreement with current observations of the Universe. During inflation, quantum fluctuations in the fields driving inflation are stretched beyond the Hubble scale, which in turn are converted to primordial curvature perturbations. In single field inflation, the comoving curvature perturbation $\zeta$ (defined further below) is conserved in superhorizon scales, which can be seen as a consequence of energy--momentum conservation \cite{Wands:2000dp}. This changes in the presence of other fields. Non--adiabatic (entropy) fluctuations are a source for the curvature perturbation and therefore, quite generally the curvature perturbation does not remain constant on superhorizon scales.

In the case of multifield inflation, the metric in field space does not have to be flat. For example, inflationary models motivated from supergravity have non--trivial fieldspace metrics quite naturally. Such inflationary models can address issues arising from the swampland program \cite{Achucarro:2018vey}. Many authors considered multifield inflationary models with a curved field manifold in the Einstein frame \cite{bib:garcia-1995,bib:dimarcofinelli-2003,bib:lalak-2007,Achucarro:2010jv,Kaiser:2013sna,bib:vandebruck-2014,vandeBruck:2015xpa,bib:longden-2016,Garcia-Saenz:2019njm,vandeBruck:2021xkm,Geller:2022nkr}.

In this work, we intend to study inflationary models with two fields, with the aim of extending the formalism of \cite{bib:dimarcofinelli-2003}. We will consider a two--field setup with a diagonal field space metric, allowing it to depend on both fields. As we will show, the amount of isocurvature perturbations produced during inflation will depend on the curvature of the field space metric, affecting the properties final curvature power spectrum, such as the spectral index and its runnings. To apply our formalism, we study three different inflationary models, each with a different field space metric. 

The paper is structured as follows. In Sec.\ref{II} we introduce the multi--field models by showing that such a curved field manifold in the Einstein frame derives from nonminimal couplings of the two fields in the Jordan one. In Sec.\ref{III} we present the homogeneous background equations and follow the conventional rotation-in-field-space approach to facilitate the interpretation of the perturbed equations of motion and Einstein equations developed in Sec.\ref{IV}. In the same section, we introduce the comoving curvature perturbation and its relation with the isocurvature mode. Furthermore, in Subsec. \ref{IVA} the slow--roll limit is studied and in Subsec. \ref{IVB} the evolution of perturbations in the super Hubble regime is considered, highlighting the role of the correlation of the perturbations. To be concrete, in Sec.\ref{V} we investigate some kinetic couplings making comparisons among them. In particular, in Subsec.\ref{VA} we deepen and complete the study of \cite{bib:longden-2016,bib:longden-universe}
by showing that even though the correlation between the two
perturbations is strong, an enlargement of the scalar spectral runnings is not present. Our conclusions can be found in Sec.\ref{VI}.

\section{Multi--field models}\label{II}
The action for the theories we consider has the form 
\begin{equation}\label{generalaction}
    {\cal S} = \int d^4 x \sqrt{-g} \left[ \frac{M_{\rm Pl}^2}{2} R - \frac{1}{2} {\cal G}_{IJ} g^{\mu\nu} \partial_\mu \phi^I \partial_\nu \phi^J - V(\phi^I)  \right],
\end{equation}
where we define $\phi^I = (\psi,\chi)$, $I,J \in \{1,2\} $ and the field space metric  ${\cal G}_{IJ}$ is a function of the two fields. Theories of this form are very well-motivated. For example, it has been shown in \cite{bib:kaiser} that starting from a Jordan frame action 
\begin{equation}
\begin{aligned}
{\cal S}_J=\int d^4x &\sqrt{-g} \bigg[ f(\psi,\chi) R  \\ 
&-\frac{1}{2}g^{\mu\nu} \partial_\mu \psi \partial_\nu \psi - \frac{1}{2}g^{\mu\nu} \partial_\mu \chi \partial_\nu \chi
- {\tilde V}(\psi,\chi)  \bigg],
\end{aligned}
\label{action}
\end{equation}
one can perform a conformal transformation of the metric $g_{\mu\nu}\rightarrow \Tilde{g}_{\mu\nu}$ such as
\begin{equation}
    \Omega^2(x) = \frac{2}{M^2_{\rm Pl}}f[(\psi(x),\chi(x))],
\end{equation}
to bring the action into the Einstein frame in which the action takes the form (\ref{generalaction}). 
Thus, calling $f = e^{h(\psi,\chi)}$ the field metric is given by 
\begin{equation}
{\cal G}_{IJ} = \frac{M^2_{\rm Pl}}{2\,f}\delta_{IJ}+\frac{3}{2}M^2_{\rm Pl}\,h_{,I}h_{,J},
\label{GIJ}
\end{equation}
where $h_{,K}=\partial h/ \partial \phi^K$. 

In general, models of the form (\ref{generalaction}) are difficult to study analytically and one has to resort to numerical methods to evaluate the power spectra at the end of inflation. However, the field space might have some symmetries in which the system simplifies considerably. Since we can perform a change of coordinates in field space and diagonalize the field metric, 
the metric from (\ref{GIJ}) can be brought in the form of
\begin{equation}
{\cal G}_{IJ} 
= \begin{pmatrix}
1 & 0 \\
0 & F(\psi,\chi) \\
\end{pmatrix}.
\label{gmatrix}
\end{equation}
One can view this as a plane polar coordinates representation of the field manifold\footnote{Needless to say, such a coordinate system might not be the best choice for studying all aspects of the model.}. Theories in which $F$ is a function of one field only (\emph{i.e.} $F = F(\psi)$, so that the field metric is shift symmetric in the $\chi$ direction) and their cosmological applications have extensively been studied in the literature \cite{bib:notari-2002,bib:dimarcofinelli-2005,bib:dimarcofinelli-2003,bib:garcia-1995,bib:lalak-2007,bib:cremonini-2011,bib:vandebruck-2014}. In this work, we relax this  requirement and allow $F$ to be a function of two fields, thereby allowing an additional coupling between the fields. Note that if $F$ is a separable function, \emph{e.g.} $F(\psi,\chi) = G(\psi)H(\chi)$, one can perform a further field redefinition of $\chi$ such that the resulting field metric takes the form (\ref{gmatrix}) but with $F=F(\psi)$. The non--canonical nature of the original field $\chi$ would then manifest itself in a change of the form of the potential $V(\psi,\chi)$. The equations derived in this paper are valid for general $F(\psi,\chi)$. To simplify the calculations and to be in line with the literature, we write $F(\psi,\chi) = e^{2b(\psi,\chi)}$, which ensures that the kinetic term does not change sign. 
The aim is to derive analytical expressions for the spectral index and its runnings, resulting from an inflationary phase.

\section{Basic equations}\label{III}
Considering the spatially flat Friedmann-Lema$\hat{\text{i}}$tre-Robertson-Walker (FRLW) Universe with line element
\begin{equation}
     ds^2=-dt^2+a(t)^2d\textbf{x}^2,
\end{equation}
where \emph{t} is the cosmic time and assuming a field space metric of the form ${\cal G}_{IJ} = \rm{diag}\left\{ 1, e^{2b(\psi,\chi)} \right\}$, we can derive the equations of motion from the action \eqref{generalaction} for the homogeneous background $\psi, \chi$. They read 
\begin{align}
    \ddot{\psi}+3H\dot{\psi}+V_{,\psi} &=e^{2b(\psi,\chi)}b_{,\psi}\dot{\chi}^2,\label{psikg}\\
    \ddot{\chi}+(3H+2b_{,\psi}\dot{\psi}+b_{,\chi}\dot{\chi})\dot{\chi} &=-e^{-2b(\psi,\chi)}V_{,\chi},
    \label{chikg}
\end{align}
whereas the Einstein equations give 
\begin{align}
    H^2&=\frac{1}{3 M_{\rm Pl}^2}\left(\frac{1}{2}\dot{\psi}^2+\frac{1}{2}e^{2b(\psi,\chi)}\dot{\chi}^2+V(\psi,\chi)\right), \label{friedman}\\
    \dot{H}&=-\frac{1}{2 M_{\rm Pl}^2}\left(\dot{\psi}^2+e^{2b(\psi,\chi)}\dot{\chi}^2\right),
\end{align}
in which $H\equiv \frac{\dot{a}}{a}$ is the Hubble rate.

Instead of working with the fields $\psi$ and $\chi$, we follow \cite{bib:gordonwands-2000,bib:dimarcofinelli-2003} and perform a field rotation and work with the fields tangential and orthogonal to the trajectory in field space. This decomposition facilitates the composition of cosmological perturbations into adiabatic and isocurvature modes, which we will study in the next section. 
In what follows we omit the $\psi$, $\chi$ dependence in $b$ to shorten the expressions. 

The \emph{adiabatic} field $\sigma$ represents the path length along the classical trajectory such that
\begin{equation}
    d\sigma=\cos\theta\,d\psi+\sin\theta e^{b}\,d\chi,
\end{equation}
whereas the (orthogonal) \emph{entropy} field $s$ is given by 
\begin{equation}
    ds = e^b\cos\theta d\chi - \sin \theta d\psi,
    \label{ds}
\end{equation}
where
\begin{equation}
    \cos\theta =\frac{\dot{\psi}}{\sqrt{\dot{\psi}^2+e^{2b}\dot{\chi}^2}},\qquad \sin\theta =\frac{e^{b}\dot{\chi}}{\sqrt{\dot{\psi}^2+e^{2b}\dot{\chi}^2}},
\end{equation}
and $\dot{\sigma}^2=\dot{\psi}^2+e^{2b}\dot{\chi}^2$. Now, adopting for simplicity the adiabatic and entropy vectors in field space as \cite{bib:lalak-2007}
\begin{equation}
\begin{aligned}
    E^I_{\sigma}&=(\cos\theta , e^{-b}\sin \theta),\\
    E^I_{s}&=(-\sin\theta , e^{-b}\cos\theta),
\end{aligned}
\end{equation}
with $I=\{\psi, \chi\}$, one can define various derivatives of the potential with respect to the adiabatic and entropy directions. They are given by
\begin{equation}
    V_{\sigma}=E^I_{\sigma}V_I, \qquad V_s=E^I_s V_I,
    \label{Vprime}
\end{equation}
whereas the second-order derivatives are
\begin{equation}
\begin{aligned}
        V_{\sigma \sigma}=E^I_{\sigma}E^J_{\sigma}&V_{IJ},\qquad V_{s s}=E^I_{s}E^J_{s}V_{IJ},\\
        &V_{\sigma s}=E^I_{\sigma}E^J_{s}V_{IJ}.
        \label{VIJ}
\end{aligned}
\end{equation}

Hence, using the definitions above
and \eqref{psikg},\eqref{chikg}, the inflationary dynamics can be described by $\sigma$ and $\theta$ instead of $\psi$ and $\chi$. The evolution equations are given by 
\begin{equation}
    \ddot{\sigma}+3H\dot{\sigma}+V_{,\sigma}=0,
\label{sigmakg}
\end{equation}
\begin{equation}
     \dot{\theta}=-\frac{V_{,s}}{\dot{\sigma}}-b_{,\psi}\dot{\sigma}\sin \theta.
\label{thetadot}
\end{equation}
Note that the form of these equations is the same as in the case for $b=b(\psi)$. The bending (\ref{thetadot}) "measures" the strength of geodesic deviation by classifying broad and sharp turn \cite{Achucarro:2010da,Cespedes:2012hu,Fumagalli:2019noh,Fumagalli:2020adf,Anguelova:2020nzl}.

\section{Perturbations}\label{IV}
We now turn our attention to the evolution of  linear cosmological perturbations in this model, focusing on scalar perturbations. Due to the vanishing anisotropic stress to the linear order
for scalar fields minimally coupled to gravity, we can write the perturbed line element in the longitudinal gauge \cite{bib:bardeen-1980} as follows
\begin{equation}
    ds^2=-(1+2\Phi)dt^2+a^2(t)(1-2\Phi)d\textbf{x}^2,
    \label{lineelementpert}
\end{equation}
in which $\Phi=\Phi(t,\textbf{x})$ represents the metric fluctuations. We consider small fluctuations $\delta \psi(t,\textbf{x})$, $\delta \chi(t,\textbf{x})$ around the background fields, so the fields, to first order, are written as 
\begin{align}
    \psi&=\psi(t)+\delta \psi(t,\textbf{x}), \\ \chi&=\chi(t)+\delta \chi(t,\textbf{x}).
\end{align}

\subsection{Adiabatic and entropy perturbations}

In Fourier space, the evolution equations for the field perturbations are given by 
\newline
\begin{equation}
    \begin{split}
    \ddot{\delta\psi}&+3H \dot{\delta\psi}+\left[\frac{k^2}{a^2}+V_{,\psi \psi}-e^{2b}(b_{,\psi \psi}+2b_{,\psi}^2)\dot{\chi}^2\right]\delta \psi\\
    &-2b_{,\psi}e^{2b}\dot{\chi}\dot{\delta\chi}+\left[V_{,\psi\chi}-e^{2b}(b_{,\psi\chi}+2b_{,\psi}b_{,\chi})\dot{\chi}^2\right]\delta \chi
    = 4 \dot{\Phi}\dot{\psi}-2V_{,\psi}\Phi,
        \end{split}
    \label{phidd}
\end{equation}
\begin{equation}
    \begin{split}
    \ddot{\delta\chi}&+(3H+2b_{,\psi}\dot{\psi}+2b_{,\chi}\dot{\chi})\,\dot{\delta\chi}+\left[\frac{k^2}{a^2}+e^{-2b}(V_{,\chi\chi}-2V_{,\chi}b_{,\chi})+b_{,\chi\chi}\dot{\chi}^2+2b_{,\chi\psi}\dot{\psi}\dot{\chi}\right]\delta{\chi}\\&+2b_{,\psi}\dot{\chi}\dot{\delta\psi} +[2b_{,\psi\psi}\dot{\psi}\dot{\chi}+e^{-2b}(V_{,\chi\psi}-2b_{,\psi}V_{,\chi})+b_{,\psi\chi}\dot{\chi}^2]\delta\psi= 4\dot{\Phi}\dot{\chi}-2\Phi e^{-2b}V_{,\chi}.
    \end{split}
    \label{chidd}
\end{equation}
\newline
The perturbed Einstein equations give 
\newline
\begin{equation}
    \begin{split}
     &3H(\dot{\Phi}+H\Phi)+\dot{H}\Phi+\frac{k^2}{a^2}\Phi \\&=-\frac{1}{2}\left[\dot{\psi}\dot{\delta\psi}+e^{2b}\dot{\chi}\dot{\delta\chi}+e^{2b}\dot{\chi}^2(b_{,\psi}\delta\psi+b_{,\chi}\delta\chi)+V_{,\chi}\delta\chi+V_{,\psi}\delta\psi\right],
    \end{split}
    \label{energy}
\end{equation}
\begin{equation}
\dot{\Phi}+H\Phi=\frac{1}{2}\left[\dot{\psi}\delta\psi+e^{2b}\dot{\chi}\delta\chi\right].
\label{momentum}
\end{equation}
\newline
A very useful quantity is the comoving curvature perturbation \cite{bib:bardeensteinard-1983,bib:lyth-1985} given in terms of the metric perturbations in the longitudinal gauge
\begin{equation}
\begin{aligned}
    \zeta&\equiv \Phi-\frac{H}{\dot{H}}(\dot{\Phi}+H\Phi) \\
    &=\Phi+H\left(\frac{\dot{\psi}\delta\psi+e^{2b}\dot{\chi}\delta\chi}{\dot{\psi}^2+e^{2b}\dot{\chi}^2}\right).
\end{aligned}
\end{equation}
Combining this equation with \eqref{energy}, \eqref{momentum}, \eqref{psikg} and \eqref{chikg} one finds an exact expression for
\begin{equation}
\begin{aligned}
\dot{\zeta}&=\frac{k^2}{a^2}\frac{H}{\dot{H}}\Phi+S\\
    &=\frac{k^2}{a^2}\frac{H}{\dot{H}}\Phi-2H\left(\frac{V_{,\psi}\dot{\psi}\dot{\chi}^2e^{2b}-V_{,\chi}\dot{\chi}\dot{\psi}^2}{(\dot{\psi}^2+e^{2b}\dot{\chi}^2)^2}\right)\left(\frac{\delta\psi}{\dot{\psi}}-\frac{\delta\chi}{\dot{\chi}}\right)\\
    &\equiv\frac{k^2}{a^2}\frac{H}{\dot{H}}\Phi-H\left[\frac{1}{2}\frac{d}{dt}\left(\frac{e^{2b}\dot{\chi}^2-\dot{\psi}^2}{e^{2b}\dot{\chi}^2+\dot{\psi}^2}\right)+\dot{C}\right]\left(\frac{\delta\psi}{\dot{\psi}}-\frac{\delta\chi}{\dot{\chi}}\right),
    \label{zetadot}
\end{aligned}
\end{equation}
where
\begin{equation}
    \dot{C}=\frac{2}{\dot{\sigma}^2}\left(b_{,\psi}\dot{\psi}\dot{\chi}^2\,e^{2b}\right)= 2b_{,\psi}\dot{\psi} \sin^2 \theta.
    \label{cdot}
\end{equation}
We note that the frictional damping \eqref{cdot} of the $\chi$ field by $\psi$ is the same obtained by \cite{bib:dimarcofinelli-2003} in the case of $b=b(\psi)$. 

After having performed the fields rotation, from \eqref{ds} it follows that perturbations with $\delta s=0$ (\emph{i.e} $s = \rm const$ along the classical trajectory) are purely adiabatic and entropy perturbations are automatically gauge-invariant \cite{bib:stewartwalker-1974}. However, in a general setting one has to define a dimensionless quantity of the total entropy fluctuation, the so--called isocurvature perturbation
\begin{equation}
    \mathcal{S}=\frac{H}{\dot{\sigma}}\delta s.
    \label{iso}
\end{equation}
In this case, it is evident the relationship between curvature and isocurvature fluctuations by using 
\begin{equation}
    \frac{\delta\psi}{\dot{\psi}}-\frac{\delta\chi}{\dot{\chi}} \equiv -\frac{ \dot{\sigma} e^{-b} }{\dot{\psi}\dot{\chi}} \delta s,
\end{equation}
so that
\begin{equation}
    S=-2\frac{V_{,s}}{\dot{\sigma}} \mathcal{S}.
\end{equation}
As expected, on super--horizon scales, \emph{i.e.} $k/a \rightarrow 0$, the isocurvature mode is a source for the curvature perturbation and the coupling through $V_{,s}$ does not vanish even when $\dot{\theta}=0$ \cite{bib:dimarcofinelli-2003}. In fact, this is directly related to the rotation of the vectors basis in the field space. In other words, $V_{,s}$ contains extra terms as we can see from \eqref{thetadot}. The assumption of having $b(\phi^I)$, with $I=\{1,2\}$, affects the feeding of the perturbations indirectly, as we will show below. 

In addition, \eqref{energy} and \eqref{momentum} can be combined to construct a gauge invariant quantity \cite{bib:bardeen-1980}, the comoving matter perturbation
\begin{equation}
    \epsilon_m \equiv \dot{\sigma}(\dot{\delta \sigma} - \dot{\sigma}\Phi)-\ddot{\sigma}\delta \sigma + 2V_{,s}\delta s,
    \label{matter}
\end{equation}
which is also related to the metric perturbation through $(k/a)^2 \Phi = -\epsilon_m/(2M_{\rm Pl}^2)$. We shall now use the new variables for the evolution of the inhomogeneities. Starting from \eqref{ds} and substituting \eqref{phidd} and \eqref{chidd}, we obtain
\begin{equation} 
\begin{aligned}
\ddot{\delta s}&+3H\dot{\delta s}+\bigg( \frac{k^2}{a^2}+V_{,ss}-\dot{\theta}^2-\dot{\sigma}^2b_{,\psi \psi}+b_{,\psi}^2 g(t)+b_{,\psi}f(t)+b_{,\chi}l(t) \bigg) \delta s \\&= -2V_s \Phi-\frac{2V_{,s}\ddot{\sigma}}{\dot{\sigma}^2}\delta \sigma +\frac{2V_{,s}}{\dot{\sigma}}\dot{\delta\sigma},
\label{dddeltas}
\end{aligned}
\end{equation}
where we have used the notation
\begin{subequations}
\begin{align}
g(t)&=-\dot{\sigma}^2(1-\sin^2{\theta}),\\
f(t)&=V_{,\psi}(1+\sin^2{\theta})+4V_{,s}\sin{\theta},\\
l(t)&=-e^{-2b}\cos^2{\theta}V_{,\chi}.\label{l(t)}
\end{align}
\end{subequations}
Additionally, highlighting the fact that at large scales $(k/a \ll H)$ entropy perturbations are decoupled from adiabatic and metric perturbations we can rewrite \eqref{dddeltas}, using \eqref{matter} and \eqref{zetadot}, as follows
\begin{equation}
\begin{aligned}
\ddot{\delta s}&+3H\dot{\delta s}+\biggl(\frac{k^2}{a^2}+V_{,ss}+3\dot{\theta}^2-\dot{\sigma}^2b_{,\psi \psi}+b_{,\psi}^2\tilde{g}(t)+b_{,\psi}\tilde{f}(t)+b_{,\chi} l(t)-4\frac{V_{,s}^2}{\dot{\sigma}^2} \biggl) \delta s\\&= \frac{2V_{,s}}{H}\dot{\zeta},
\label{dddeltasreduced}
\end{aligned}
\end{equation}
where now
\begin{subequations}
\begin{align}
    \tilde{g}(t)&=-\dot{\sigma}^2(1+3\sin^2{\theta}),\\
    \tilde{f}(t)&=V_{,\psi}(1+\sin^2\theta)-4V_{,s}\sin\theta. 
\end{align}
\end{subequations}

It is interesting to note that (\ref{zetadot}) takes the same form as in the case in which $b=b(\psi)$ only. The additional dependence of the field metric on $\chi$ enters only indirectly via the evolution of the fields. In the case of (\ref{dddeltasreduced}), there is an explicit dependence on $b_{,\chi}$, changing the effective mass of the entropy perturbation $\delta s$. 

Before we apply these equations to the slow--roll case and consider specific models, it is illuminating to relate our results to results found in the literature. While a full discussion is beyond the aim of this paper, it would be interesting to compare the formalism and results of in this paper to the more geometric methods put forward in other works, such as e.g. \cite{Kaiser:2013sna,Achucarro:2010jv,Cespedes:2012hu}. Such work would provide a dictionary between these formalisms, which could lead to more insight. For the work here, let us compare the effective mass of the entropy field found above to the effective mass of the entropy perturbation derived in \cite{Cespedes:2012hu} which takes the form 

\begin{equation}
    m_{s(\rm eff)}^2 = V_{;ss}-(H\eta_{\perp})^2+\epsilon H^2 M_{\rm Pl}^2 R_{\rm fs},
    \label{meff1}
\end{equation}
where the dimensionless parameter $\eta_{\perp}\equiv-E_s^IV_{,I}/(H\dot{\sigma})$ measures the deviation of the background trajectory from a field space geodesic \cite{GrootNibbelink:2000vx,bib:groot-2001}, $V_{;ss}=E_{s}^IE_{s}^J(V_{,IJ}-\Gamma^K_{IJ}V_{,K})$ is the projection of the covariant Hessian of the potential along the entropic direction ($\Gamma^K_{IJ}$ denote the Christoffel symbols for the field space) and $R_{\rm fs}$ is the Ricci scalar of the field space. 
In our case, 
\begin{equation}
    \eta_{\perp}=\frac{\dot{\theta}}{H}+b_{\psi}\frac{\dot{\sigma}}{H}\sin \theta,
\end{equation}
 \begin{equation}
    V_{;ss}=V_{,ss}+2b_{,\psi}e^{-b}\sin \theta \cos\theta V_{,\chi}+b_{,\psi}(1-\sin^2 \theta)V_{,\psi}-b_{,\chi}e^{-2b}\cos^2{\theta}V_{,\chi},
\end{equation}
and 
\begin{equation}
R_{\rm fs} = -2\left(b_{,\psi}^2 + b_{,\psi \psi} \right).    
\end{equation}
Note that the Ricci scalar depends only on the derivatives with respect to $\psi$ but not the second field $\chi$. Collecting all terms and going on large scales, we find that $m^2_{s(\rm eff)}$ is given by $$ m^2_{s(\rm eff)} = V_{,ss}+3\dot{\theta}^2-\dot{\sigma}^2b_{,\psi \psi}+b_{,\psi}^2\tilde{g}(t)+b_{,\psi}\tilde{f}(t)+b_{,\chi} l(t).$$ The term involving the derivative $b_{,\chi}$ has its origin in the derivative $V_{;ss}$. Note that the sign of the $b_{,\chi}$--term (see (\ref{l(t)})) depends on the product $b_{,\chi}V_{,\chi}$. It should be noted that we can relate (\ref{l(t)}) to the velocity of the second field in the slow--roll approximation, as we will show below. 

When the kinetic energy density $\epsilon H^2 M_{\rm Pl}^2$ grows enough during inflation, in case of a negative $R_{\rm fs}$, it can turn the effective mass of the entropy field from positive to negative values, leading to geometrical destabilization \cite{Renaux-Petel:2015mga}. In the following sections, we will avoid geometrical destabilization. In the models we consider, turns of the field trajectory happen towards the end of inflation, at which entropic fluctuations have sufficiently decayed, as we will show in Sec.\ref{V}.

Having found the exact equations for the evolution of curvature and entropy perturbations, we are now
considering the large-wavelength limit for slow-rolling fields.

\subsection{Slow--roll approximation}\label{IVA}
A common approach to studying the inflationary phase is the slow--roll approximation in which the following are satisfied
\begin{equation}
\begin{aligned}
    \epsilon=&-\frac{\dot{H}}{H^2} \ll 1 \quad \Rightarrow \quad \mathcal{G}_{IJ}\dot{\phi}^I\dot{\phi}^J \ll V(\phi^I),\\
    \epsilon_1=&\frac{\dot{\epsilon}}{H \epsilon} \ll 1 \quad \Rightarrow \quad 2 \dot{\phi}_I D_t(\dot{\phi}^I)\ll H \;\mathcal{G}_{IJ}\dot{\phi}^I\dot{\phi}^J.\\\label{epsilons}
    \end{aligned}
\end{equation}
In fact, we are assuming below that both fields are slowly rolling and therefore not considering models such as hyperinflation, angular inflation, and side-tracked inflation \cite{Achucarro:2015rfa,Christodoulidis:2018qdw,Linde:2018hmx,Brown:2017osf,Achucarro:2019pux,Mizuno:2017idt,Bjorkmo:2019aev}. 

Under the slow-roll conditions, the equation of motion \eqref{psikg}, \eqref{chikg} and the Friedmann equation \eqref{friedman} can be simplified as
\begin{equation}
\begin{aligned}
    \dot{\sigma}\cos \theta=\dot{\psi} \simeq-&\frac{V_{,\psi}}{3H}, \quad \dot{\sigma}\sin \theta e^{-b}=\dot{\chi}\simeq-\frac{V_{,\chi}}{3H}e^{-2b},\\
    &H^2(\psi,\chi) \simeq \frac{V(\psi,\chi)}{3 M_{\rm Pl}^2}.
    \label{fieldexpansion}
\end{aligned}
\end{equation}
Moreover, the background slow--roll solution is 
\begin{equation}
-\frac{\dot{\theta}}{H}
\simeq  \eta_{,\sigma s}+\frac{\dot{\sigma}}{H}b_{,\psi}\sin \theta \cos^2 \theta +\frac{\dot{\sigma}}{H}b_{,\chi} e^{-b}\sin^2 \theta \cos \theta,
\label{dthetaslow}
\end{equation}
\begin{equation}
\frac{\ddot{\sigma}}{H\dot{\sigma}}
\simeq \epsilon-\eta_{,\sigma \sigma}-\frac{\dot{\sigma}}{H}b_{,\psi}\sin^2\theta \cos\theta - \frac{\dot{\sigma}}{H}b_{,\chi}e^{-b}\sin^3\theta,
\label{ddsigmaslow}
\end{equation}
in which
\begin{equation}
    \eta_{,IJ}=\frac{V_{,IJ}}{3H^2}.
\label{etaIJ}
\end{equation}

In this regime, the power spectrum $\mathcal{P}_{\zeta}$ of the curvature perturbation at horizon crossing ($k  = a H$) is given by \cite{bib:garriga-1999,bib:langlois-2008}
\begin{equation}
    \mathcal{P^*_{\zeta}}\simeq \frac{H^2}{8\pi^2 \epsilon}.
\end{equation}
Introducing a weak scale dependence in the primordial spectrum modelled by the running of the scalar tilt and its running, it is possible to show that the growth due to the kinetic coupling is already present and plays a role at the horizon crossing in terms of slow--roll parameters. Hence, the spectral index reads as
\begin{equation}
    n_* = \frac{\rm d \ln \mathcal{P_{\zeta}}}{H \rm dt} \biggl|_*= \frac{1}{H}\biggl[ \frac{\epsilon}{H^2}\biggl( \frac{2H\dot{H}\epsilon - \dot{\epsilon}H^2}{\epsilon^2} \biggl) \biggl]\biggl|_* \\
    = -2\epsilon_*-\epsilon_{1*}.
\end{equation}
Now, considering \eqref{epsilons} and making use of 
\begin{equation}
    \dot{\epsilon}=2H \epsilon (2 \epsilon-\eta_{,\sigma \sigma}-\sqrt{2\epsilon}b_{,\psi}\sin^2\theta \cos\theta-e^{-b}\sqrt{2\epsilon}b_{,\chi}\sin^3\theta),
\end{equation}
we arrive at  
\begin{equation}
    n_*=-6\epsilon + 2\eta_{,\sigma \sigma} + 2\sqrt{2\epsilon} b_{,\psi}\sin^2\theta \cos \theta+2e^{-b}\sqrt{2\epsilon} b_{,\chi}\sin^3\theta,
    \label{nstar}
\end{equation}
where in the limit of a flat field metric, \eqref{nstar} is equal to the expression presented in \cite{bib:dimarcofinelli-2005}.
 We relate the higher-order running
\begin{equation}
    \alpha_* \equiv \frac{\text{d} n}{H \rm dt}\biggl|_* = -2\frac{\dot{\epsilon_*}}{H_*}-\frac{\dot{\epsilon}_{1*}}{H_*},
\end{equation}
which quantifies the rate of change of $n$ per Hubble time.
We find 
\begin{equation}
\begin{aligned}
    \alpha_*=&-24 \epsilon^2-4 \eta _{,\sigma s}^2+16 \epsilon \eta _{,\sigma \sigma }-2\bar{\xi}_1^2 \sin ^4 \theta e^{-2b}-2 \bar{\xi}_1 \xi_1 \sin ^3\theta \cos \theta e^{-b}-2 \bar{\xi}_1  \xi_1\sin ^3\theta \cos ^3\theta e^{-b}\\&-4 \bar{\xi}_1  \xi_1 \sin ^5\theta \cos \theta e^{-b}-6 \bar{\xi}_1 \eta _{,\sigma \sigma }\sin ^3\theta e^{-b}-14 \bar{\xi}_1  \eta _{,\sigma s} \sin ^2\theta \cos \theta e^{-b}-2 \bar{\xi}_1 ^2 \sin ^6\theta e^{-2 b}\\&+6 \bar{\xi}_1 ^2 \sin ^4\theta \cos ^2\theta e^{-2 b}+16 \bar{\xi}_1  \epsilon \sin ^3\theta e^{-b}+2 \bar{\xi}_{12}  \sin ^3\theta \cos \theta e^{-b}+2 \bar{\xi}_2 \sin ^4\theta e^{-2 b}\\& + 2 \xi_2 \sin^2 \theta \cos^2 \theta +2 \xi_1  \eta _{,\sigma s} \sin ^3\theta-12 \xi_1 \eta _{,\sigma s} \sin \theta  \cos ^2\theta-6 \xi_1  \eta _{,\sigma \sigma } \sin ^2\theta \cos \theta\\&-4 \xi_1^2 \sin ^2\theta  \cos ^4\theta +16 \xi_1 \epsilon \sin ^2\theta \cos \theta-2 \alpha _{,\sigma \sigma \sigma }, \label{alpha}
\end{aligned}
\end{equation}
where we follow \cite{bib:lalak-2007} in the definition of new slow--roll parameters given by 
\begin{equation}
\begin{aligned}\label{xi-parameter}
    \xi_1 &= \sqrt{2 \epsilon} b_{,\psi}, \quad 
    \xi_2 = 2 \epsilon b_{,\psi \psi},\\
    \bar{\xi}_1&= \sqrt{2\epsilon}b_{,\chi},\quad
     \bar{\xi}_2= 2\epsilon b_{,\chi \chi},\quad \bar{\xi}_{12}= 2\epsilon b_{,\chi \psi},
\end{aligned}
\end{equation}
and $\alpha_{,IJK}=V_{,\sigma}V_{,IJK}/V^2$.

An expression for the running of the running 
\begin{equation}
    \beta_* \equiv \frac{\text{d}  \alpha}{H \rm dt}\biggl|_*,
\end{equation}
can be found in the same way. However, the expression is cumbersome, we therefore refrain to write it down here. The calculations of $n_{*}$, $\alpha_{*}$ and $\beta_*$ in the next section have been done using the expressions above as well as our ana\-ly\-tical expressions and we found excellent agreement between these two methods.

\subsection{Super--horizon scales}\label{IVB}
As it is well known in models with multiple fields and easy to see from \eqref{zetadot}, in the super--Hubble regime curvature perturbations are sourced solely by the entropy field. Hence, the evolution of curvature and isocurvature perturbation can be written in terms of slow-roll parameters
\begin{equation}
\begin{aligned}
    \dot{\zeta}&\simeq-2\frac{V_{,s}}{\dot{\sigma}}\mathcal{S}=AH\mathcal{S}, \\
    \dot{\mathcal{S}}&\simeq\frac{H}{\dot{\sigma}}\dot{\delta}s+\left(\frac{\dot{H}}{H^2}-\frac{\ddot{\sigma}}{H\dot{\sigma}}\right)\frac{H^2}{\dot{\sigma}}\delta s= BH\mathcal{S},
    \label{Sdot}
    \end{aligned}
\end{equation}
in which $A, B$ are  time--dependent dimensionless functions obtained substituting \eqref{dddeltasreduced}, \eqref{dthetaslow}, \eqref{ddsigmaslow} and read as
\begin{equation}
\begin{aligned}
    A&=-2\eta_{,\sigma s}+2 \xi_1\sin^3 \theta-2\bar{\xi}_1e^{-b}\sin^2 \theta \cos \theta \\
    &=2\frac{\dot\theta}{H} + 2 \xi_1 \sin\theta, \\
    B&= -\eta_{,ss}+\eta_{,\sigma \sigma}-2\epsilon+\xi_1\cos\theta(1+2\sin^2 \theta)+\bar{\xi}_1 e^{-b} \sin\theta(2 \sin^2 \theta -1)+\frac{\xi_1^2}{3}+\frac{\xi_2}{3}.
    \label{AB1}
    \end{aligned}
\end{equation}
To derive these equations we used the fact that on superhorizon scales \cite{bib:groot-2001,bib:gordonwands-2000}
\begin{equation}
    |\ddot{\delta s}|\ll 3H |\dot{\delta s}|, \quad |\ddot{\delta \sigma}|\ll 3H |\dot{\delta \sigma}|.
\end{equation}
To calculate the curvature perturbation spectrum at the end of inflation, we have to take into account the effect of isocurvature modes. In order to do this, we integrate \eqref{Sdot} over time and relate curvature and entropy perturbations at Hubble crossing to those at some later time through the transfer matrix \cite{bib:wands-2002}
\[
\begin{pmatrix}
\zeta\\
\mathcal{S}
\end{pmatrix}
=
\begin{pmatrix}
1 & \mathcal{T}_{\zeta \mathcal{S}}\\
0 & \mathcal{T}_{\mathcal{S}\mathcal{S}}
\end{pmatrix}
\begin{pmatrix}
\zeta\\
\mathcal{S}
\end{pmatrix}_*,
\]
where
\begin{equation}
    \begin{aligned}
        \mathcal{T}_{\zeta \mathcal{S}}(t_*,t)&=\int^t_{t_*} A(t')H(t')\mathcal{T}_{\mathcal{S} \mathcal{S}}(t_*,t')dt',\\
        \mathcal{T}_{\mathcal{S} \mathcal{S}}(t_*,t')&=\text{exp}\left(\int^{t'}_{t_{*}} B(t'')H(t'')dt''\right).
    \label{transf}
        \end{aligned}
\end{equation}
Although $A$ and $B$, defined in (\ref{AB1}), are constant during slow--roll, they could undergo a significant variation towards the end of inflation. Hence, we do not assume their constancy when performing the integrations in (\ref{transf}).

We relate the curvature power spectrum at the end of inflation to the one at horizon crossing via the transfer angle, defined by 
\begin{equation}
    \mathcal{P_{\zeta}}=(1+\mathcal{T}_{\zeta\mathcal{S}}^2) \mathcal{P^*_{\zeta}}\equiv \frac{\mathcal{P^*_{\zeta}}}{\cos^2\Theta}.
    \label{pr}
\end{equation}
The transfer angle depends on the parameters \eqref{AB1}, and therefore also on the geometry of the curved field metric. The mix of curvature and isocurvature perturbations can be uncorrelated, but some amount of correlation arises between them if the trajectory in field space is curved during inflation. 

Furthermore, one can derive expressions for the spectral index and runnings at the end of inflation in a general two--field model \cite{bib:longden-2016}
\begin{equation}
\begin{aligned}
n_s &\simeq n_*-2\sin{\Theta}(A_* \cos{\Theta}+B_*\sin{\Theta}),\\
\alpha_s &\simeq\alpha_*+2\cos{\Theta}
 (A_*\cos{\Theta}+B_*\sin{\Theta})\times (A_*\cos{2\Theta}+B_*\sin{2\Theta}),\\
\beta_s&\simeq \beta_*-2\cos{\Theta}(A_*\cos{\Theta}+B_*\sin{\Theta})\times (B_*\cos{2\Theta}-A_*\sin{2\Theta})\\&\times (A_*+2A_*\cos{2\Theta}+B_*\sin{2\Theta}).
\label{index}
\end{aligned}
\end{equation}
As a side observation and a consequence of what we explained above, the transfer function contains the integral of $A$ and $B$, which are quite small on average from horizon crossing to the end of inflation, 
because they are of the order of the slow--roll parameter. The analytical expressions for the spectral indices at the end of inflation above contain a combination of trigonometric functions, which could lead to an increase or decrease in the values of the spectral indices between horizon crossing and the end of inflation. This also results in an increase of the amplitude (\ref{pr}), which mainly depends on the value of the effective entropy mass from the Hubble crossing until the time of the turn (unlike $\mathcal{P}_{\zeta}^*$).
Moreover, looking at (\ref{Sdot}), the curvature perturbation is fed by the entropy perturbation $\delta s$, which is decaying as we find the parameter $B$ to be negative. Hence, these quantities determine whether the power spectrum amplitude, spectral index, or the runnings at the end of inflation deviate from the initial values at horizon crossing.

\section{Investigation of some kinetic couplings}\label{V}
In this section, we apply our formalism to three inflationary models. To be concrete, we keep potential fixed to be that of two interacting massive scalar fields
\begin{equation}
    V = \frac{1}{2} m_\psi^2 \psi^2 + \frac{1}{2}m_\chi^2 \chi^2 + g^2\psi^2 \chi^2,
    \label{potential}
\end{equation}
in which $m_{I}$ ($I=\{\psi,\chi\}$) denotes the masses of the fields and $g$ a coupling constant. We first study the simplest case in which the kinetic coupling is only a function of one field $\psi$. We then consider a couple of choices for the kinetic coupling, which depends on both fields. 

\subsection{$b(\psi) = - c\frac{\psi}{M_{\rm Pl}}$}\label{VA}
This choice of coupling has been extensively studied in the literature, see \cite{bib:notari-2002,bib:dimarcofinelli-2005,bib:cremonini-2011,bib:garcia-1995,bib:dimarcofinelli-2003,bib:lalak-2007,bib:vandebruck-2014}, and we include it here for completeness.

\begin{figure}[h!]
\centering
    \includegraphics[width=12cm]{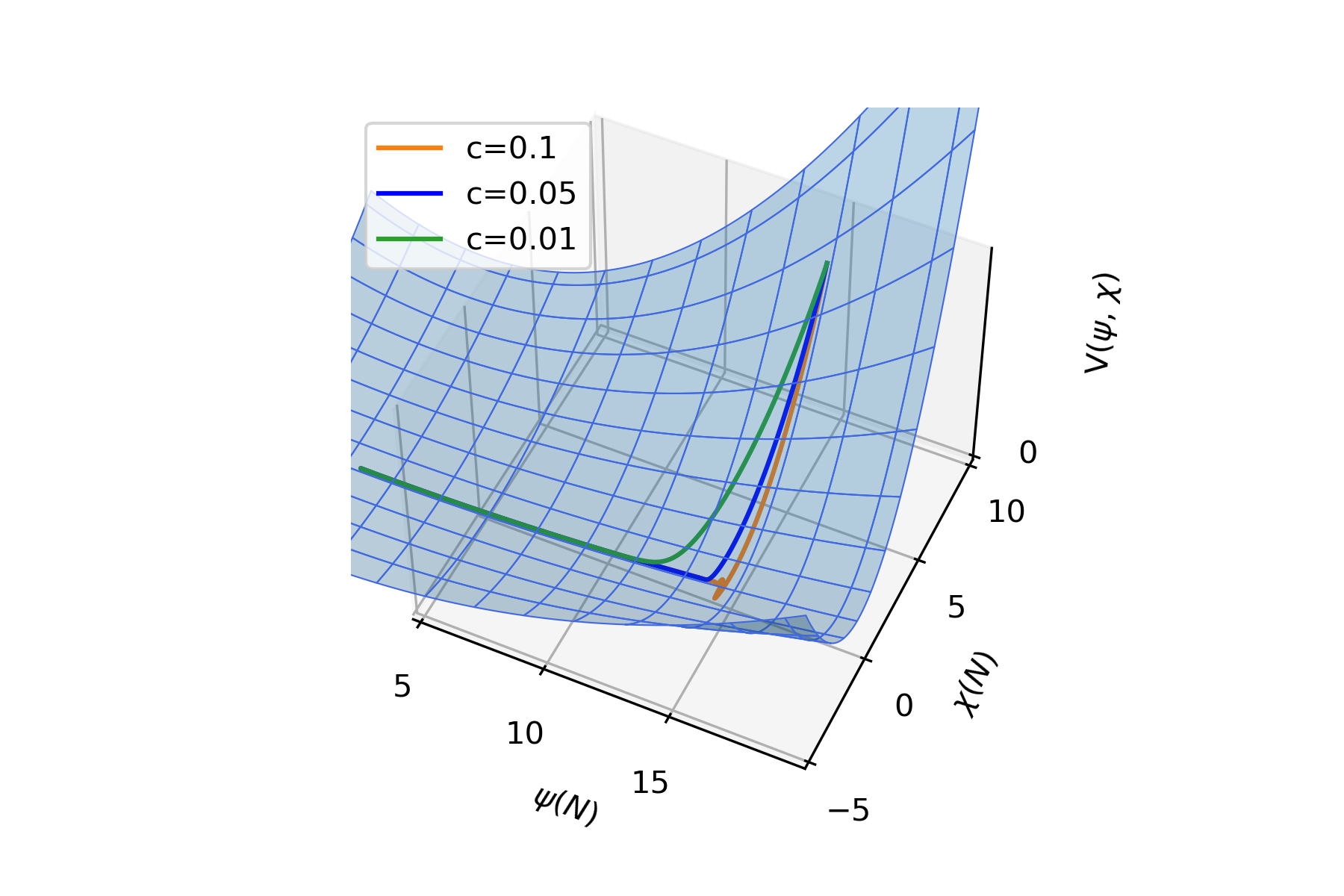}
    \caption{Fields trajectories on the potential $V(\psi,\chi)$ for different values of the kinetic coupling constant $c$ (the values for $\psi$ and $\chi$ are in Planck units). It has been used $m_{\psi}=3 \cdot 10^{-6}$ $\text{M}_{\text{Pl}}$, $m_{\chi}=6 \cdot 10^{-6}$ $\text{M}_{\text{Pl}}$, $g= 10^{-6}$, $N=55$ e--folds.  }  
    \label{fig:trajectoriescpositive}
\end{figure}

\begin{figure}[h!]
\centering
    \includegraphics[width=12cm]{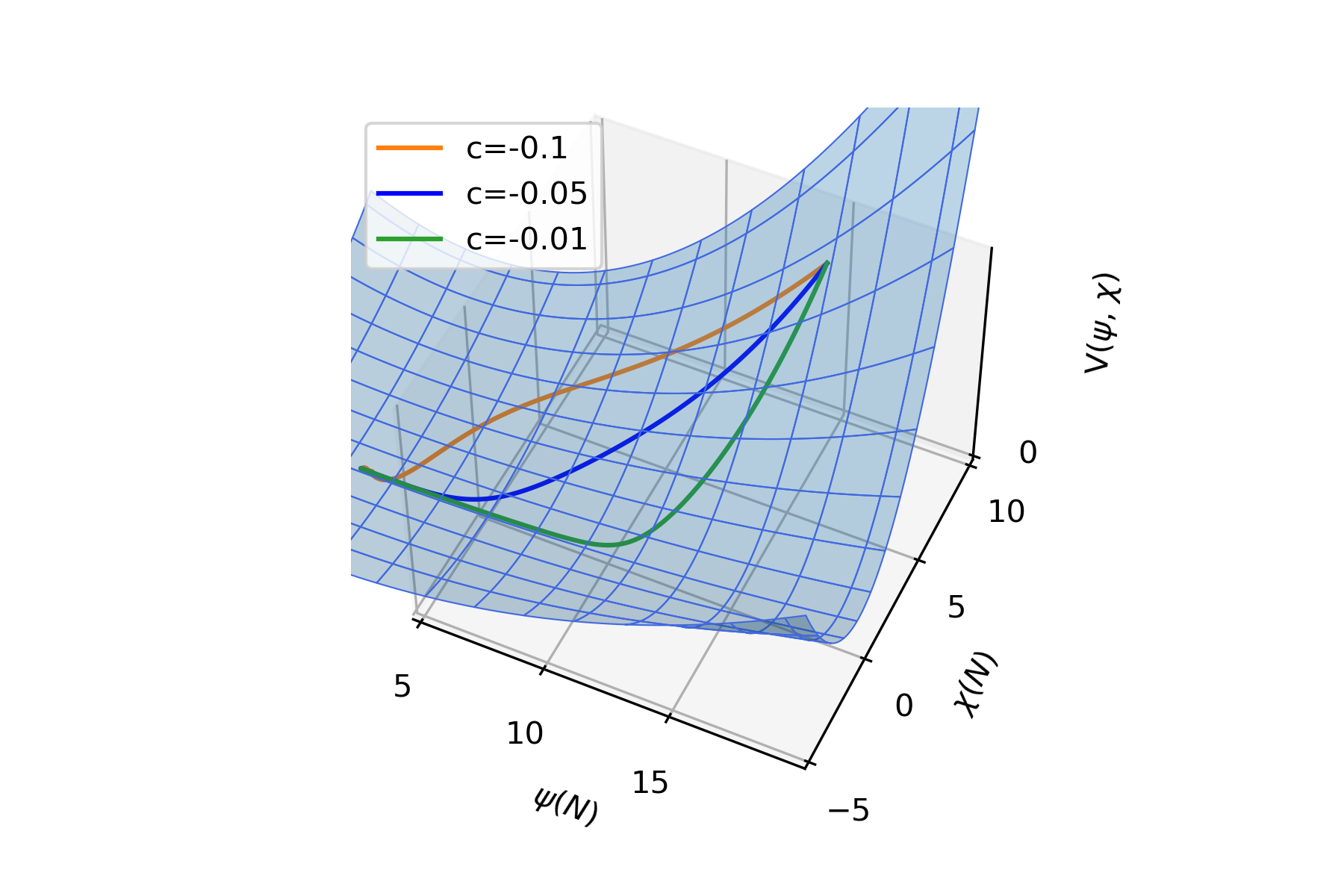}
    \caption{Fields trajectories on the potential $V(\psi,\chi)$ for different values of the kinetic coupling constant $c$ (the values for $\psi$ and $\chi$ are in Planck units). We have set here $m_{\psi}=3 \cdot 10^{-6}$ $\text{M}_{\text{Pl}}$ , $m_{\chi}=6 \cdot 10^{-6}$ $\text{M}_{\text{Pl}}$, $g= 10^{-6}$, $N=55$ e--folds.}    
    \label{fig:trajectoriescnegative}
\end{figure}

If we assume that $\psi$ is positive during inflation and is rolling towards the minimum of the potential during inflation, the sign of $c$ determines whether the $\chi$ field is driven quickly to zero, resulting in an inflationary period driven by one field only, or whether it slowly rolls towards the minimum, jointly with $\psi$. In Figs.  \ref{fig:trajectoriescpositive} and \ref{fig:trajectoriescnegative} we show the field trajectories for both cases. The case of positive $c$ is not of interest here, because inflation is driven essentially by $\psi$ only. As we can see from Fig \ref{fig:trajectoriescnegative}, the more negative $c$ is, the longer it takes for $\chi$ to reach $\chi = 0$. 
The results of our numerical calculation for the power spectrum give 
\begin{equation}
 \begin{split}
    \mathcal{P}_{\zeta}^* & \simeq 1.98 \times 10^{-9},\\
    n_*&= 0.958,\\
    \alpha_*&= -7.87 \times 10^{-4}, \\
    \beta_*&= -2.45 \times 10^{-5}
  \end{split}
  \qquad
  \begin{split}
    \mathcal{P}_{\zeta}& \simeq 2.08 \times 10^{-9},\\
    n_s&= 0.956,\\
    \alpha_s&= -8.00 \times 10^{-4}, \\
    \beta_s&= -2.44 \times 10^{-5},  \\
  \end{split}
\end{equation}
where the masses are chosen as in Fig. \ref{fig:trajectoriescnegative} and $c=-0.1$. We calculated $\beta$ via numerical integration.

Firstly, we note that the amplitude of the power spectrum increases slightly from horizon crossing to the end of inflation, meaning that there is a small amount of entropy perturbations affecting the curvature perturbations after horizon crossing (see (\ref{Sdot})). Indeed, the values of $A$ and $B$ at horizon crossing are $A_*=-7 \times 10^{-3}$, $B_*=-0.02$ and we find  $|\mathcal{T}_{\zeta S}|= 0.23$. 
This choice of the free parameters implies that for the inflationary model under consideration, the interplay between entropy and adiabatic perturbations is not of great relevance. We do not observe the inversion of the standard hierarchy $|\alpha_s| > |\beta_s|$ as reported in \cite{bib:longden-2016}. 
Note that from  \eqref{AB1}, $A$ depends on the turning rate $\dot\theta/H$. Indeed, the more negative $c$ is, the more $\chi$ gets stuck, reducing the amount of entropy perturbations produced during inflation. In \cite{bib:longden-2016}, it is pointed out that at the leading order behavior of $\Theta$, a positive amplification of $\beta_s$ is possible if $B_*$ is negative and large. However, considering this regime one is implicitly taking into account small values of $A$ and $B$.

\subsection{$b(\psi,\chi) = -c\frac{\psi\chi}{M_{\rm Pl}^2}$}

\begin{figure}[h!]
    \centering
    \includegraphics[width=14cm]{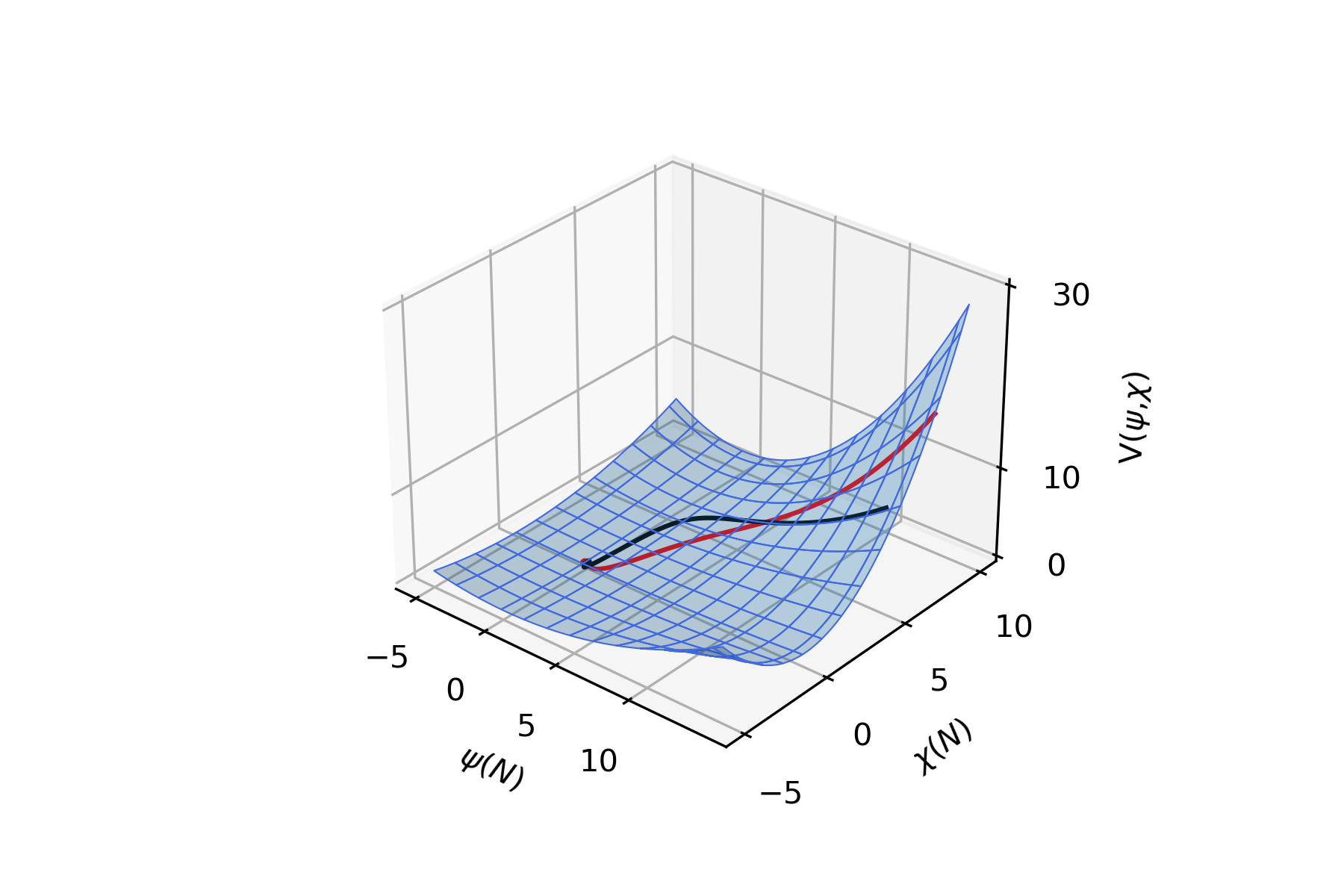}
    \caption{Representation of the fields trajectories on the potential by considering $b(\psi)$, red line, and $b(\psi$,$\chi)$ in black. The parameter has been set $c=-0.05$ in both cases, the parameter $g$, $m_\psi$ and $m_\chi$ are chosen as in Fig. \ref{fig:trajectoriescnegative}. The values for $\psi$ and $\chi$ are in Planck units. }
    \label{fig:trajectories}
\end{figure}

We now turn our attention to a field space metric, in which the function $b(\psi,\chi)$ is a function of both fields and cannot be reduced further by means of a field--redefinition. In Fig.\ref{fig:trajectories} we show two trajectories, one for the case of $b=b(\psi)$ (red curve) and $b=b(\psi,\chi)$ (black curve), with the same choice of values for $c$, $m_\psi$, $m_\chi$ and $g$. We fix $\psi_{\rm ini}$ but vary $\chi_{\rm ini}$ so that the duration of inflation is the same. For the black trajectory, the $\chi$ field starts rolling only after $\psi$ is near the minimum ($\psi=0$). This behaviour will cause $\eta_{,ss}$ to be relatively large, implying the decay of isocurvature mode. Moreover, by decreasing $c$ further, $\chi$ is stuck for longer, reducing the relevance of isocurvature modes. For the inflationary model under consideration, we obtain
\begin{equation}
 \begin{split}
    \mathcal{P}_{\zeta}^* &\simeq 2.14 \times 10^{-9} ,\\
    n_*&= 0.953 ,\\
    \alpha_*&= -1.01 \times 10^{-3}, \\
    \beta_*&= -4.33 \times 10^{-5} ,
  \end{split}
  \qquad
  \begin{split}
    \mathcal{P}_{\zeta} &\simeq 2.15 \times 10^{-9},\\
    n_s&= 0.954,\\
    \alpha_s&= -1.00 \times 10^{-3}, \\
    \beta_s&= -4.26 \times 10^{-5},  \\
  \end{split}
\end{equation}
where, as expected, $A_*= -1 \times 10^{-3}$, $B_*=-0.05$, $|\mathcal{T}_{\zeta S}|= 0.04$, $|\Theta| = 0.04$, highlighting the fact that entropy perturbations do not play an essential role in this case. Indeed, entropy fluctuations have substantially decayed until the rapid turn occurs towards the end of inflation.

\subsection{$2b(\psi,\chi) = \ln\left[\frac{\left(a \psi + b \chi \right)^2}{M_{\rm Pl}^2} \right] $}

In the following, we choose another kinetic coupling function, such that a further simplification of the kinetic terms is not possible. The term multiplying the kinetic part of $\chi$ is now $(a\psi + b\chi)^2$, which can, depending on the sign of $a$ or $b$, become zero. In what follows, we avoid this situation, by choosing $a$ and $b$ having the same sign. In our calculations, we set $a=b=0.2$ and the parameter $g$, $m_\psi$ and $m_\chi$ are chosen as in Fig. \ref{fig:trajectoriescnegative}.
We obtain 
\begin{equation}
 \begin{split}
    \mathcal{P}_{\zeta}^* &\simeq 3.14 \times 10^{-10},\\
    n_*&= 0.970,\\
    \alpha_*&=  -7.05 \times 10^{-4}, \\
    \beta_*&= -3.26 \times 10^{-5},
  \end{split}
  \qquad
  \begin{split}
    \mathcal{P}_{\zeta}&\simeq 2.01 \times 10^{-9},\\
    n_s&=  0.961,\\
    \alpha_s&= -8.63 \times 10^{-4}, \\
    \beta_s&= -3.61 \times 10^{-5}.  \\
  \end{split}
\end{equation}
We find $|\mathcal{T}_{\zeta S}|= 2.37$, $|\Theta| = 1.72$, $A_*=-0.04$, $B_*=-0.01$ and $|\eta_{,\sigma s}|\gg \{|\eta_{,s s}|,|\eta_{,\sigma \sigma}|\}$ (implying $|A|>|B|$). In this model, the interplay between curvature and isocurvature perturbations is significant ($|\mathcal{T}_{\zeta S}|>1$). The spectral index only slightly changes during inflation after horizon crossing, while the runnings $\alpha_s$ and $\beta_s$ change considerably but stay small after horizon crossing. As a final remark, we note that the choice of $a$ and $b$ in this model determines whether the inflationary trajectory in field space is curved or not. In our numerical analysis, we find for example that fixing $b$ but varying $a$ determines whether the trajectory is curved or more straight. Increasing $a$ in this case results in trajectories which are more curved. As a result, entropy perturbations play more of a role for smaller values of $a$. 

\section{Conclusions}\label{VI}
In this paper we have studied two--field inflationary scenarios, allowing for a non--trivial field metric. We extended the work of \cite{bib:garcia-1995,bib:dimarcofinelli-2003,bib:lalak-2007} by relaxing the assumptions of a shift--symmetry in the direction of the field we denote by $\chi$. Our aim was to find out how far the production and evolution of entropy and curvature perturbations is affected by reducing the symmetry of the field space. Our first main result is that the evolution equation for the curvature perturbation $\zeta$ is unchanged (our equations (\ref{zetadot}) and (\ref{cdot})). The evolution equation for $\zeta$ contains the derivative of $b$ with respect to the field $\psi$, but not $\chi$. The effect of the dependence on $\chi$ in this equation enters only indirectly by the (background) evolution of $\psi$ and $\chi$. On the other hand, the reduced symmetry does affect the effective mass of the entropy perturbation $\delta s$. As it can be seen from equation (\ref{dddeltasreduced}), depending on the sign of the product $V_{,\chi} b_{,\chi}$, the (effective) mass can be enhanced or reduced. The derivative of $b$ with respect to the $\chi$--field only appears explicitly in the evolution equation for the entropy perturbation \eqref{dddeltasreduced}. Of course, the evolution of the {\it background} fields is affected by the reduced symmetry in field space. 

We have then considered the slow--roll limit of the equations, finding expressions for the spectral index $n_s$ \eqref{nstar} and its running $\alpha_s$ \eqref{alpha} at horizon crossing. The resulting quantities at the end of inflation can also be obtained, assuming slow--roll inflation throughout (see \eqref{index}). Note that the parameter $A$ and $B$ do depend on the derivatives of both fields, via the slow--roll parameter $\xi_1$, $\xi_2$, $\bar\xi_1$, $\bar\xi_2$, defined in \eqref{xi-parameter}. 

The formalism developed is widely applicable, also in e.g. bouncing cosmologies. We have studied three phenomenological inflationary models in Section \ref{V}. Each model differs by the choice of the coupling function $b(\psi,\chi)$ but has the same potential $V(\phi,\chi)$. Two of the models discussed have non--trivial field metrics depending on both fields. In one of the models (model 3), we observe a significant effect of entropy perturbations on the adiabatic modes after horizon crossing. The outcome arises from the fact that in this case, a smoother function $f(\psi,\chi)$ has been selected. Moreover, the kinetic term for $\chi$ reads $(a\psi + b\chi)^2(\partial \chi)^2$. For the choice of parameters, the relative change of $n_s$ after horizon crossing is small (the relative change of $n_s$ from horizon crossing to the end of inflation is less than a percent), but the runnings $\alpha_s$ and $\beta_s$ change more (18 percent and 10 percent, respectively). However, in none of the models considered we observe an inversion of the hierarchy $|n_s-1| > |\alpha_s| > |\beta_s|$. 

In future work, we introduce an innovative sampling algorithm, capable of efficiently exploring the parameter space and comparing multifield inflationary scenarios to observations, such as anisotropies in the CMB. This allows us to identify regions in parameter space for which theoretical predictions align with observational data \cite{Giare:2023kiv}.

\section*{Acknowledgments}
We are grateful to Ana Achucarro for very valuable comments on our work. MDA thanks Chiara Cecchini for pointing out a typo in \eqref{AB1}.
MDA is supported by an EPSRC studentship. CvdB is supported (in part) by the Lancaster–Manchester–Sheffield Consortium for Fundamental Physics under STFC grant: ST/T001038/1.

\printbibliography

@article{bib:kaiser,
  title = {Conformal transformations with multiple scalar fields},
  author = {Kaiser, David I.},
  journal = {Phys. Rev. D},
  volume = {81},
  issue = {8},
  pages = {084044},
  numpages = {8},
  year = {2010},
  month = {Apr},
  publisher = {American Physical Society},
  doi = {10.1103/PhysRevD.81.084044},
  url = {https://link.aps.org/doi/10.1103/PhysRevD.81.084044}
}

@article{Geller:2022nkr,
    author = "Geller, Sarah R. and Qin, Wenzer and McDonough, Evan and Kaiser, David I.",
    title = "{Primordial black holes from multifield inflation with nonminimal couplings}",
    eprint = "2205.04471",
    archivePrefix = "arXiv",
    primaryClass = "hep-th",
    reportNumber = "MIT-CTP/5426",
    doi = "10.1103/PhysRevD.106.063535",
    journal = "Phys. Rev. D",
    volume = "106",
    number = "6",
    pages = "063535",
    year = "2022"
}

@article{Kaiser:2013sna,
    author = "Kaiser, David I. and Sfakianakis, Evangelos I.",
    title = "{Multifield Inflation after Planck: The Case for Nonminimal Couplings}",
    eprint = "1304.0363",
    archivePrefix = "arXiv",
    primaryClass = "astro-ph.CO",
    reportNumber = "PREPRINT-MIT-CTP-4451",
    doi = "10.1103/PhysRevLett.112.011302",
    journal = "Phys. Rev. Lett.",
    volume = "112",
    number = "1",
    pages = "011302",
    year = "2014"
}

@article{Achucarro:2018vey,
    author = "Ach\`{u}carro, Ana and Palma, Gonzalo A.",
    title = "{The string swampland constraints require multi-field inflation}",
    eprint = "1807.04390",
    archivePrefix = "arXiv",
    primaryClass = "hep-th",
    doi = "10.1088/1475-7516/2019/02/041",
    journal = "JCAP",
    volume = "02",
    pages = "041",
    year = "2019"
}

@article{bib:notari-2002,
title = {Isocurvature perturbations in the Ekpyrotic Universe},
journal = {Nuclear Physics B},
volume = {644},
number = {1},
pages = {371-382},
year = {2002},
issn = {0550-3213},
doi = {https://doi.org/10.1016/S0550-3213(02)00765-4},
url = {https://www.sciencedirect.com/science/article/pii/S0550321302007654},
author = {A. Notari and A. Riotto}
}

@article{bib:dimarcofinelli-2003,
  title = {Adiabatic and isocurvature perturbations for multifield generalized Einstein models},
  author = {Di Marco, F. and Finelli, F. and Brandenberger, R.},
  journal = {Phys. Rev. D},
  volume = {67},
  issue = {6},
  pages = {063512},
  numpages = {11},
  year = {2003},
  month = {Mar},
  publisher = {American Physical Society},
  doi = {10.1103/PhysRevD.67.063512},
  url = {https://link.aps.org/doi/10.1103/PhysRevD.67.063512}
}

@article{bib:dimarcofinelli-2005,
  title = {Slow-roll inflation for generalized two-field Lagrangians},
  author = {Di Marco, Fabrizio and Finelli, Fabio},
  journal = {Phys. Rev. D},
  volume = {71},
  issue = {12},
  pages = {123502},
  numpages = {9},
  year = {2005},
  month = {Jun},
  publisher = {American Physical Society},
  doi = {10.1103/PhysRevD.71.123502},
  url = {https://link.aps.org/doi/10.1103/PhysRevD.71.123502}
}

@article{bib:gordonwands-2000,
  title = {Adiabatic and entropy perturbations from inflation},
  author = {Gordon, Christopher and Wands, David and Bassett, Bruce A. and Maartens, Roy},
  journal = {Phys. Rev. D},
  volume = {63},
  issue = {2},
  pages = {023506},
  numpages = {11},
  year = {2000},
  month = {Dec},
  publisher = {American Physical Society},
  doi = {10.1103/PhysRevD.63.023506},
  url = {https://link.aps.org/doi/10.1103/PhysRevD.63.023506}
}

@ARTICLE{bib:stewartwalker-1974,
       author = {{Stewart}, J.~M. and {Walker}, M.},
        title = "{Perturbations of space-times in general relativity}",
      journal = {Proceedings of the Royal Society of London Series A},
         year = 1974,
        month = oct,
       volume = {341},
       number = {1624},
        pages = {49-74},
          doi = {10.1098/rspa.1974.0172},
       adsurl = {https://ui.adsabs.harvard.edu/abs/1974RSPSA.341...49S}
}

@article{bib:bardeen-1980,
  title = {Gauge-invariant cosmological perturbations},
  author = {Bardeen, James M.},
  journal = {Phys. Rev. D},
  volume = {22},
  issue = {8},
  pages = {1882--1905},
  numpages = {0},
  year = {1980},
  month = {Oct},
  publisher = {American Physical Society},
  doi = {10.1103/PhysRevD.22.1882},
  url = {https://link.aps.org/doi/10.1103/PhysRevD.22.1882}
}

@article{bib:bardeensteinard-1983,
  title = {Spontaneous creation of almost scale-free density perturbations in an inflationary universe},
  author = {Bardeen, James M. and Steinhardt, Paul J. and Turner, Michael S.},
  journal = {Phys. Rev. D},
  volume = {28},
  issue = {4},
  pages = {679--693},
  numpages = {0},
  year = {1983},
  month = {Aug},
  publisher = {American Physical Society},
  doi = {10.1103/PhysRevD.28.679},
  url = {https://link.aps.org/doi/10.1103/PhysRevD.28.679}
}

@article{bib:lyth-1985,
  title = {Large-scale energy-density perturbations and inflation},
  author = {Lyth, D. H.},
  journal = {Phys. Rev. D},
  volume = {31},
  issue = {8},
  pages = {1792--1798},
  numpages = {0},
  year = {1985},
  month = {Apr},
  publisher = {American Physical Society},
  doi = {10.1103/PhysRevD.31.1792},
  url = {https://link.aps.org/doi/10.1103/PhysRevD.31.1792}
}

@article{Guth:1980zm,
    author = "Guth, Alan H.",
    editor = "Fang, Li-Zhi and Ruffini, R.",
    title = "{The Inflationary Universe: A Possible Solution to the Horizon and Flatness Problems}",
    reportNumber = "SLAC-PUB-2576",
    doi = "10.1103/PhysRevD.23.347",
    journal = "Phys. Rev. D",
    volume = "23",
    pages = "347--356",
    year = "1981"
}

@inproceedings{Baumann:2009ds,
    author = "Baumann, Daniel",
    title = "{Inflation}",
    booktitle = "{Theoretical Advanced Study Institute in Elementary Particle Physics}: {Physics of the Large and the Small}",
    eprint = "0907.5424",
    archivePrefix = "arXiv",
    primaryClass = "hep-th",
    reportNumber = "TASI-2009",
    doi = "10.1142/9789814327183_0010",
    pages = "523--686",
    year = "2011"
}

@article{Wands:2000dp,
    author = "Wands, David and Malik, Karim A. and Lyth, David H. and Liddle, Andrew R.",
    title = "{A New approach to the evolution of cosmological perturbations on large scales}",
    eprint = "astro-ph/0003278",
    archivePrefix = "arXiv",
    doi = "10.1103/PhysRevD.62.043527",
    journal = "Phys. Rev. D",
    volume = "62",
    pages = "043527",
    year = "2000"
}

@article{Burgess:2013sla,
    author = "Burgess, C. P. and Cicoli, M. and Quevedo, F.",
    title = "{String Inflation After Planck 2013}",
    eprint = "1306.3512",
    archivePrefix = "arXiv",
    primaryClass = "hep-th",
    reportNumber = "DAMTP-2013-32",
    doi = "10.1088/1475-7516/2013/11/003",
    journal = "JCAP",
    volume = "11",
    pages = "003",
    year = "2013"
}

@article{Burgess:2011fa,
    author = "Burgess, C. P. and McAllister, Liam",
    title = "{Challenges for String Cosmology}",
    eprint = "1108.2660",
    archivePrefix = "arXiv",
    primaryClass = "hep-th",
    doi = "10.1088/0264-9381/28/20/204002",
    journal = "Class. Quant. Grav.",
    volume = "28",
    pages = "204002",
    year = "2011"
}

@article{McAllister:2014mpa,
    author = "McAllister, Liam and Silverstein, Eva and Westphal, Alexander and Wrase, Timm",
    title = "{The Powers of Monodromy}",
    eprint = "1405.3652",
    archivePrefix = "arXiv",
    primaryClass = "hep-th",
    reportNumber = "SU/ITP-14/13, SLAC-PUB-15962, DESY-14-078, SU-ITP-14-13",
    doi = "10.1007/JHEP09(2014)123",
    journal = "JHEP",
    volume = "09",
    pages = "123",
    year = "2014"
}

@article{Baumann:2007ah,
    author = "Baumann, Daniel and Dymarsky, Anatoly and Klebanov, Igor R. and McAllister, Liam",
    title = "{Towards an Explicit Model of D-brane Inflation}",
    eprint = "0706.0360",
    archivePrefix = "arXiv",
    primaryClass = "hep-th",
    reportNumber = "PUPT-2237, ITEP-TH-22-07",
    doi = "10.1088/1475-7516/2008/01/024",
    journal = "JCAP",
    volume = "01",
    pages = "024",
    year = "2008"
}

@article{Firouzjahi:2003zy,
    author = "Firouzjahi, Hassan and Tye, S. H. Henry",
    title = "{Closer towards inflation in string theory}",
    eprint = "hep-th/0312020",
    archivePrefix = "arXiv",
    doi = "10.1016/j.physletb.2004.01.022",
    journal = "Phys. Lett. B",
    volume = "584",
    pages = "147--154",
    year = "2004"
}

@article{Kachru:2003sx,
    author = "Kachru, Shamit and Kallosh, Renata and Linde, Andrei D. and Maldacena, Juan Martin and McAllister, Liam P. and Trivedi, Sandip P.",
    title = "{Towards inflation in string theory}",
    eprint = "hep-th/0308055",
    archivePrefix = "arXiv",
    reportNumber = "SLAC-PUB-9669, SU-ITP-03-18, TIFR-TH-03-06",
    doi = "10.1088/1475-7516/2003/10/013",
    journal = "JCAP",
    volume = "10",
    pages = "013",
    year = "2003"
}

@article{Starobinsky:1980te,
    author = "Starobinsky, Alexei A.",
    editor = "Khalatnikov, I. M. and Mineev, V. P.",
    title = "{A New Type of Isotropic Cosmological Models Without Singularity}",
    doi = "10.1016/0370-2693(80)90670-X",
    journal = "Phys. Lett. B",
    volume = "91",
    pages = "99--102",
    year = "1980"
}

@article{Linde:1983gd,
    author = "Linde, Andrei D.",
    title = "{Chaotic Inflation}",
    doi = "10.1016/0370-2693(83)90837-7",
    journal = "Phys. Lett. B",
    volume = "129",
    pages = "177--181",
    year = "1983"
}

@article{Albrecht:1982wi,
    author = "Albrecht, Andreas and Steinhardt, Paul J.",
    editor = "Fang, Li-Zhi and Ruffini, R.",
    title = "{Cosmology for Grand Unified Theories with Radiatively Induced Symmetry Breaking}",
    reportNumber = "UPR-0185T",
    doi = "10.1103/PhysRevLett.48.1220",
    journal = "Phys. Rev. Lett.",
    volume = "48",
    pages = "1220--1223",
    year = "1982"
}

@article{bib:lalak-2007,
doi = {10.1088/1475-7516/2007/07/014},
url = {https://dx.doi.org/10.1088/1475-7516/2007/07/014},
year = {2007},
month = {jul},
publisher = {},
volume = {2007},
number = {07},
pages = {014},
author = {Z Lalak and D Langlois and S Pokorski and K Turzyński},
title = {Curvature and isocurvature perturbations in two-field inflation},
journal = {Journal of Cosmology and Astroparticle Physics}
}

@article{bib:cremonini-2011,
doi = {10.1088/1475-7516/2011/03/016},
url = {https://dx.doi.org/10.1088/1475-7516/2011/03/016},
year = {2011},
month = {mar},
publisher = {},
volume = {2011},
number = {03},
pages = {016},
author = {Sera Cremonini and  Zygmunt Lalak and  Krzysztof Turzyński},
title = {Strongly coupled perturbations in two-field inflationary models},
journal = {Journal of Cosmology and Astroparticle Physics}
}

@article{bib:garriga-1999,
    author = "Garriga, Jaume and Mukhanov, Viatcheslav F.",
    title = "{Perturbations in k-inflation}",
    eprint = "hep-th/9904176",
    archivePrefix = "arXiv",
    reportNumber = "UAB-FT-466",
    doi = "10.1016/S0370-2693(99)00602-4",
    journal = "Phys. Lett. B",
    volume = "458",
    pages = "219--225",
    year = "1999"
}

@article{bib:langlois-2008,
    author = "Langlois, David and Renaux-Petel, Sebastien and Steer, Daniele A. and Tanaka, Takahiro",
    title = "{Primordial perturbations and non-Gaussianities in DBI and general multi-field inflation}",
    eprint = "0806.0336",
    archivePrefix = "arXiv",
    primaryClass = "hep-th",
    reportNumber = "CERN-PH-TH-2008-098",
    doi = "10.1103/PhysRevD.78.063523",
    journal = "Phys. Rev. D",
    volume = "78",
    pages = "063523",
    year = "2008"
}

@article{bib:groot-2001,
    author = "Groot Nibbelink, S. and van Tent, B. J. W.",
    title = "{Scalar perturbations during multiple field slow-roll inflation}",
    eprint = "hep-ph/0107272",
    archivePrefix = "arXiv",
    reportNumber = "SPIN-2001-19, ITP-UU-01-27",
    doi = "10.1088/0264-9381/19/4/302",
    journal = "Class. Quant. Grav.",
    volume = "19",
    pages = "613--640",
    year = "2002"
}

@article{bib:wands-2002,
    author = "Wands, David and Bartolo, Nicola and Matarrese, Sabino and Riotto, Antonio",
    title = "{An Observational test of two-field inflation}",
    eprint = "astro-ph/0205253",
    archivePrefix = "arXiv",
    reportNumber = "PU-ICG-02-06",
    doi = "10.1103/PhysRevD.66.043520",
    journal = "Phys. Rev. D",
    volume = "66",
    pages = "043520",
    year = "2002"
}

@article{bib:longden-2016,
  title = {Running of the running and entropy perturbations during inflation},
  author = {van de Bruck, Carsten and Longden, Chris},
  journal = {Phys. Rev. D},
  volume = {94},
  issue = {2},
  pages = {021301},
  numpages = {6},
  year = {2016},
  month = {Jul},
  publisher = {American Physical Society},
  doi = {10.1103/PhysRevD.94.021301},
  url = {https://link.aps.org/doi/10.1103/PhysRevD.94.021301}
}

@Article{bib:longden-universe,
AUTHOR = {Longden, Chris},
TITLE = {Non-Standard Hierarchies of the Runnings of the Spectral Index in Inﬂation},
JOURNAL = {Universe},
VOLUME = {3},
YEAR = {2017},
NUMBER = {1},
ARTICLE-NUMBER = {17},
URL = {https://www.mdpi.com/2218-1997/3/1/17},
ISSN = {2218-1997},
DOI = {10.3390/universe3010017}
}

@article{Giare:2023kiv,
    author = "Giar\`e, William and De Angelis, Mariaveronica and van de Bruck, Carsten and Di Valentino, Eleonora",
    title = "{Tracking the Multifield Dynamics with Cosmological Data: A Monte Carlo approach}",
    eprint = "2306.12414",
    archivePrefix = "arXiv",
    primaryClass = "astro-ph.CO",
    month = "6",
    year = "2023"
}

@article{Obied:2018sgi,
    author = "Obied, Georges and Ooguri, Hirosi and Spodyneiko, Lev and Vafa, Cumrun",
    title = "{De Sitter Space and the Swampland}",
    eprint = "1806.08362",
    archivePrefix = "arXiv",
    primaryClass = "hep-th",
    reportNumber = "CALT-TH-2018-020, IPMU18-0100",
    month = "6",
    year = "2018"
}

@article{Borde:2001nh,
    author = "Borde, Arvind and Guth, Alan H. and Vilenkin, Alexander",
    title = "{Inflationary space-times are incompletein past directions}",
    eprint = "gr-qc/0110012",
    archivePrefix = "arXiv",
    reportNumber = "MIT-CTP-3183",
    doi = "10.1103/PhysRevLett.90.151301",
    journal = "Phys. Rev. Lett.",
    volume = "90",
    pages = "151301",
    year = "2003"
}

@article{Agrawal:2018own,
    author = "Agrawal, Prateek and Obied, Georges and Steinhardt, Paul J. and Vafa, Cumrun",
    title = "{On the Cosmological Implications of the String Swampland}",
    eprint = "1806.09718",
    archivePrefix = "arXiv",
    primaryClass = "hep-th",
    doi = "10.1016/j.physletb.2018.07.040",
    journal = "Phys. Lett. B",
    volume = "784",
    pages = "271--276",
    year = "2018"
}

@book{Baumann:2014nda,
    author = "Baumann, Daniel and McAllister, Liam",
    title = "{Inflation and String Theory}",
    eprint = "1404.2601",
    archivePrefix = "arXiv",
    primaryClass = "hep-th",
    doi = "10.1017/CBO9781316105733",
    isbn = "978-1-107-08969-3, 978-1-316-23718-2",
    publisher = "Cambridge University Press",
    series = "Cambridge Monographs on Mathematical Physics",
    month = "5",
    year = "2015"
}

@article{bib:garcia-1995,
  title = {Constraints from inflation on scalar-tensor gravity theories},
  author = {Garc\'{i}a-Bellido, Juan and Wands, David},
  journal = {Phys. Rev. D},
  volume = {52},
  issue = {12},
  pages = {6739--6749},
  numpages = {0},
  year = {1995},
  month = {Dec},
  publisher = {American Physical Society},
  doi = {10.1103/PhysRevD.52.6739},
  url = {https://link.aps.org/doi/10.1103/PhysRevD.52.6739}
}

@article{Achucarro:2010jv,
    author = "Achucarro, Ana and Gong, Jinn-Ouk and Hardeman, Sjoerd and Palma, Gonzalo A. and Patil, Subodh P.",
    title = "{Mass hierarchies and non-decoupling in multi-scalar field dynamics}",
    eprint = "1005.3848",
    archivePrefix = "arXiv",
    primaryClass = "hep-th",
    reportNumber = "CPHT-RR-039.0510, LPTENS-10-20",
    doi = "10.1103/PhysRevD.84.043502",
    journal = "Phys. Rev. D",
    volume = "84",
    pages = "043502",
    year = "2011"
}

@article{Garcia-Saenz:2019njm,
    author = "Garcia-Saenz, Sebastian and Pinol, Lucas and Renaux-Petel, S\'{e}bastien",
    title = "{Revisiting non-Gaussianity in multifield inflation with curved field space}",
    eprint = "1907.10403",
    archivePrefix = "arXiv",
    primaryClass = "hep-th",
    doi = "10.1007/JHEP01(2020)073",
    journal = "JHEP",
    volume = "01",
    pages = "073",
    year = "2020"
}

@article{vandeBruck:2015xpa,
    author = "van de Bruck, Carsten and Paduraru, Laura Elena",
    title = "{Simplest extension of Starobinsky inflation}",
    eprint = "1505.01727",
    archivePrefix = "arXiv",
    primaryClass = "hep-th",
    doi = "10.1103/PhysRevD.92.083513",
    journal = "Phys. Rev. D",
    volume = "92",
    pages = "083513",
    year = "2015"
}

@article{vandeBruck:2021xkm,
    author = "van de Bruck, Carsten and Daniel, Richard",
    title = "{Inflation and scale-invariant R$^2$ gravity}",
    eprint = "2102.11719",
    archivePrefix = "arXiv",
    primaryClass = "gr-qc",
    doi = "10.1103/PhysRevD.103.123506",
    journal = "Phys. Rev. D",
    volume = "103",
    number = "12",
    pages = "123506",
    year = "2021"
}

@article{bib:vandebruck-2014,
doi = {10.1088/1475-7516/2014/08/024},
url = {https://dx.doi.org/10.1088/1475-7516/2014/08/024},
year = {2014},
month = {aug},
publisher = {},
volume = {2014},
number = {08},
pages = {024},
author = {Carsten van de Bruck and Mathew Robinson},
title = {Power spectra beyond the slow roll approximation in theories with non-canonical kinetic terms},
journal = {Journal of Cosmology and Astroparticle Physics}
}

@article{Renaux-Petel:2015mga,
    author = "Renaux-Petel, S\'ebastien and Turzy\'nski, Krzysztof",
    title = "{Geometrical Destabilization of Inflation}",
    eprint = "1510.01281",
    archivePrefix = "arXiv",
    primaryClass = "astro-ph.CO",
    doi = "10.1103/PhysRevLett.117.141301",
    journal = "Phys. Rev. Lett.",
    volume = "117",
    number = "14",
    pages = "141301",
    year = "2016"
}

@article{Achucarro:2010da,
    author = "Achucarro, Ana and Gong, Jinn-Ouk and Hardeman, Sjoerd and Palma, Gonzalo A. and Patil, Subodh P.",
    title = "{Features of heavy physics in the CMB power spectrum}",
    eprint = "1010.3693",
    archivePrefix = "arXiv",
    primaryClass = "hep-ph",
    reportNumber = "LPTENS-10-36, CPHT-RR-080.0910",
    doi = "10.1088/1475-7516/2011/01/030",
    journal = "JCAP",
    volume = "01",
    pages = "030",
    year = "2011"
}

@article{GrootNibbelink:2000vx,
    author = "Groot Nibbelink, S. and van Tent, B. J. W.",
    title = "{Density perturbations arising from multiple field slow roll inflation}",
    eprint = "hep-ph/0011325",
    archivePrefix = "arXiv",
    reportNumber = "SPIN-2000-31, ITP-UU-00-35, NIKHEF-00-038",
    month = "11",
    year = "2000"
}

@article{Achucarro:2015rfa,
    author = "Ach\'ucarro, Ana and Atal, Vicente and Welling, Yvette",
    title = "{On the viability of $m^2\phi^2$ and natural inflation}",
    eprint = "1503.07486",
    archivePrefix = "arXiv",
    primaryClass = "astro-ph.CO",
    doi = "10.1088/1475-7516/2015/07/008",
    journal = "JCAP",
    volume = "07",
    pages = "008",
    year = "2015"
}

@article{Christodoulidis:2018qdw,
    author = "Christodoulidis, Perseas and Roest, Diederik and Sfakianakis, Evangelos I.",
    title = "{Angular inflation in multi-field $\alpha$-attractors}",
    eprint = "1803.09841",
    archivePrefix = "arXiv",
    primaryClass = "hep-th",
    doi = "10.1088/1475-7516/2019/11/002",
    journal = "JCAP",
    volume = "11",
    pages = "002",
    year = "2019"
}

@article{Linde:2018hmx,
    author = "Linde, Andrei and Wang, Dong-Gang and Welling, Yvette and Yamada, Yusuke and Ach\'ucarro, Ana",
    title = "{Hypernatural inflation}",
    eprint = "1803.09911",
    archivePrefix = "arXiv",
    primaryClass = "hep-th",
    doi = "10.1088/1475-7516/2018/07/035",
    journal = "JCAP",
    volume = "07",
    pages = "035",
    year = "2018"
}

@article{Brown:2017osf,
    author = "Brown, Adam R.",
    title = "{Hyperbolic Inflation}",
    eprint = "1705.03023",
    archivePrefix = "arXiv",
    primaryClass = "hep-th",
    doi = "10.1103/PhysRevLett.121.251601",
    journal = "Phys. Rev. Lett.",
    volume = "121",
    number = "25",
    pages = "251601",
    year = "2018"
}

@article{Achucarro:2019pux,
    author = "Ach\'ucarro, Ana and Copeland, Edmund J. and Iarygina, Oksana and Palma, Gonzalo A. and Wang, Dong-Gang and Welling, Yvette",
    title = "{Shift-symmetric orbital inflation: Single field or multifield?}",
    eprint = "1901.03657",
    archivePrefix = "arXiv",
    primaryClass = "astro-ph.CO",
    reportNumber = "DESY-19-015",
    doi = "10.1103/PhysRevD.102.021302",
    journal = "Phys. Rev. D",
    volume = "102",
    number = "2",
    pages = "021302",
    year = "2020"
}

@article{Mizuno:2017idt,
    author = "Mizuno, Shuntaro and Mukohyama, Shinji",
    title = "{Primordial perturbations from inflation with a hyperbolic field-space}",
    eprint = "1707.05125",
    archivePrefix = "arXiv",
    primaryClass = "hep-th",
    reportNumber = "YITP-17-73, IPMU17-0099",
    doi = "10.1103/PhysRevD.96.103533",
    journal = "Phys. Rev. D",
    volume = "96",
    number = "10",
    pages = "103533",
    year = "2017"
}

@article{Bjorkmo:2019aev,
    author = "Bjorkmo, Theodor and Marsh, M. C. David",
    title = "{Hyperinflation generalised: from its attractor mechanism to its tension with the \textquoteleft{}swampland conditions\textquoteright{}}",
    eprint = "1901.08603",
    archivePrefix = "arXiv",
    primaryClass = "hep-th",
    doi = "10.1007/JHEP04(2019)172",
    journal = "JHEP",
    volume = "04",
    pages = "172",
    year = "2019"
}

@article{Cespedes:2012hu,
    author = "Cespedes, Sebastian and Atal, Vicente and Palma, Gonzalo A.",
    title = "{On the importance of heavy fields during inflation}",
    eprint = "1201.4848",
    archivePrefix = "arXiv",
    primaryClass = "hep-th",
    doi = "10.1088/1475-7516/2012/05/008",
    journal = "JCAP",
    volume = "05",
    pages = "008",
    year = "2012"
}

@article{Anguelova:2020nzl,
    author = "Anguelova, Lilia",
    title = "{On Primordial Black Holes from Rapid Turns in Two-field Models}",
    eprint = "2012.03705",
    archivePrefix = "arXiv",
    primaryClass = "hep-th",
    doi = "10.1088/1475-7516/2021/06/004",
    journal = "JCAP",
    volume = "06",
    pages = "004",
    year = "2021"
}

@article{Fumagalli:2019noh,
    author = "Fumagalli, Jacopo and Garcia-Saenz, Sebastian and Pinol, Lucas and Renaux-Petel, S\'ebastien and Ronayne, John",
    title = "{Hyper-Non-Gaussianities in Inflation with Strongly Nongeodesic Motion}",
    eprint = "1902.03221",
    archivePrefix = "arXiv",
    primaryClass = "hep-th",
    doi = "10.1103/PhysRevLett.123.201302",
    journal = "Phys. Rev. Lett.",
    volume = "123",
    number = "20",
    pages = "201302",
    year = "2019"
}

@article{Fumagalli:2020adf,
    author = "Fumagalli, Jacopo and Renaux-Petel, S\'ebastien and Ronayne, John W. and Witkowski, Lukas T.",
    title = "{Turning in the landscape: A new mechanism for generating primordial black holes}",
    eprint = "2004.08369",
    archivePrefix = "arXiv",
    primaryClass = "hep-th",
    doi = "10.1016/j.physletb.2023.137921",
    journal = "Phys. Lett. B",
    volume = "841",
    pages = "137921",
    year = "2023"
}
\end{document}